\documentclass[12pt,english]{article}

\usepackage{graphics,graphpap,mathdots}
\usepackage[english]{babel}
\usepackage{amsfonts}
\usepackage{hyperref}

\newcommand{\bea}{\begin{eqnarray}}
\newcommand{\eea}{\end{eqnarray}}
\newcommand{\beq}{\begin{equation}}
\newcommand{\eeq}{\end{equation}}
\newcommand{\be}{\begin{equation}}
\newcommand{\ee}{\end{equation}}
\newcommand{\sst}{\scriptstyle}

\def\nn{\nonumber}
\def\nl{\nonumber\\}
\def\barr#1{\begin{array}{#1}}
\def\earr{\end{array}}
\def\disp{\displaystyle}

\begin{document}
\begin{center}
\textbf{\LARGE A closed expression for the UV-divergent parts of one-loop tensor integrals
 in dimensional regularization }
\end{center}
\vspace{10mm}
\begin{center}
\large
G. Sulyok$^1$
\end{center}
\begin{center}
\textit{
$^1$ Institute of Atomic and Subatomic Physics, Vienna University of Technology }

\textit{
Stadionallee 2, 1020 Vienna, Austria}
\end{center}
\vspace{30mm}
\large \textbf{Abstract}
\newline
\newline
\normalsize Starting from the general definition of a one-loop tensor N-point function, we use its Feynman parametrization to calculate the UV-divergent part of an arbitrary tensor coefficient in the framework of dimensional regularization. In contrast to existing recursion schemes, we are able to present a general analytic result in closed form that enables direct determination of the UV-divergent part of any one-loop tensor N-point coefficient independent from UV-divergent parts of other one-loop tensor N-point coefficients. Simplified formulas and explicit expressions are presented for A-, B-, C-, D-, E-, and F-functions.

\section{Introduction}
Quantum field theory provides a perturbation series expansion which allows the calculation of physical observables in particles processes up to an arbitrary accuracy \cite{ryder,peskin}. Due to high precision measurements of decay rates and cross sections at collider experiments, present calculations include at least quantum corrections of first order. The evaluation of radiative corrections contains integrals over the inner momentum. These loop integrals have been investigated and classified by t'Hooft, Veltman and Passarino
\cite{'tHooft:1978xw,PaVe}, partly elaborating an idea of Brown and Feynman \cite{Brown:1952eu}. Their works focus on loop integrals with up to four internal propagators. Nevertheless, their methods can be extended leading to a compact notation for all one loop tensor N-point functions and coefficients, presented in
\cite{Denner:1993kt} and exhaustively used in \cite{Denner:2005nn}.

The divergent behaviour of these loop integrals necessitates the renormalization procedure rendering all physical observables finite. To quantify the divergences of the various loop integrals a regularization scheme is needed. Dimensional regularization has become the standard method to deal with ultraviolet (UV-) divergences \cite{'tHooft:1972fi,Bollini:1972bi,Ashmore:1972uj}. In current literature, one can find extensive tables of UV-divergent parts of specific tensor coefficients \cite{Denner:2005nn,Denner:2002ii} and generally valid recursive evaluation schemes \cite{Denner:2005nn,Feng_recursive}, but no closed analytic formula is given yet. Using Feynman parametrization, we are able to provide such an expression for the UV-divergent part of an arbitrary one-loop tensor N-point coefficient.

Although the number of physically relevant tensor functions is limited by dimensional considerations or renormalizability, tensor coefficients of arbitrary high ranks can appear as mathematical by-product as, for example, in the reduction schemes for one loop tensor integrals in case of vanishing Gram determinants \cite{Denner:2005nn}. These schemes provide recursive formulas that converge for infinite iteration steps. To get high order approximations for a certain tensor coefficient, one needs lower order expressions for tensor coefficients of a higher rank. In particular, the divergent parts of these high rank tensor coefficients are necessary as well. Since it is not possible to decide a priori which approximation order suffices for a specific process, a generally valid formula for divergent parts represents a useful tool.

We have organized this work as follows: First, all occurring quantities are defined and a compact notation is introduced. Then, we outline the key steps of the calculation and present the general result. In addition, we provide simplified formulas for A-, B-, C-, and D-functions and and in the appendix, we give explicit expressions for some B-, C-, D-, E-, and F- tensor coefficients.

\section{Definitions and notation}
The general form of a one-loop tensor N-point integral reads
\be\label{def}
T^{N, \ \mu_1 \dots \mu_{P}}(p_1,\dots,p_{N-1},m_0,\dots,m_{N-1}) :=
\frac{(2\pi \mu)^{4-D}}{i \pi^2} \int {\rm d^D q}
\frac{q^{\mu_1}\dots q^{\mu_{P}}}{N_0 N_1 \dots N_{N-1}}
\ee
with denominators
\be
N_k = (q + p_k)^2 - m_k^2 + i\eta, \quad (p_0 \equiv 0) 
\ee 
where $i\eta$ ($\eta > 0$) denotes an infinitesimally small imaginary part, $\mu$ is a mass parameter and $D$ the non-integer space-time dimension defined as $D=4-\varepsilon$. Following common convention, we abbreviate $T^1=A$, $T^2=B$, $T^3=C$, $T^4=D$, etc. For the decomposition of the tensor integral in its Lorentz-covariant structures we use the same notation as in
\cite{Denner:2005nn}
\bea\label{gentensor} 
T^{N,\mu_1\ldots\mu_P}
&=&\sum_{i_1,\ldots,i_P=1}^{N-1} p^{\mu_1}_{i_1}\ldots p^{\mu_P}_{i_P}
T^N_{i_1\ldots i_P}
+ \sum_{i_3,\ldots,i_{P}=1}^{N-1}
\{g p\ldots p\}^{\mu_1\ldots\mu_P}_{i_3\ldots i_P} T^N_{00i_3\ldots i_{P}}
\nl&&{}
+ \sum_{i_5,\ldots,i_{P}=1}^{N-1}
\{g g p\ldots p\}^{\mu_1\ldots\mu_P}_{i_5 \ldots i_P} T^N_{0000i_5\ldots i_{P}}
+\ldots
\nl&&{}+
\left\{
\barr{cc}\disp
\sum_{i_{P}=1}^{N-1}
\{g \ldots g p\}_{i_P}^{\mu_1\ldots\mu_P}
T^N_{\underbrace{\sst0\ldots 0}_{P-1} i_{P}},
& \mbox{ for } P\ \mathrm{odd,} \\
\{g \ldots g \}^{\mu_1\ldots\mu_P}
T^N_{\underbrace{\sst0\ldots 0}_{P}},
& \mbox{ for } P\ \mathrm{even.}
\earr\right.
\eea
The curly brackets stand for symmetrization with respect to Lorentz indices in such a way that all non-equivalent permutations of the Lorentz indices on metric tensors $g$ and a
generic momentum $p$ contribute with weight one. In covariants with $n$ momenta $p_{i_j}^{\mu_j}$ $(j=1,\dots,n$) only one representative out of the $n!$ permutations of the indices $i_j$ is
kept, e.g.
\bea
\{gg\}^{\mu\nu\rho\sigma} &=& g^{\mu\nu}g^{\rho\sigma}+g^{\nu\rho}g^{\mu\sigma}
+g^{\rho\mu}g^{\nu\sigma},
\nn\\[.3em]
\{g p\}_{i_1}^{\mu\nu\rho} &=&
g^{\mu\nu}p_{i_1}^{\rho}+g^{\nu\rho}p_{i_1}^{\mu}+g^{\rho\mu}p_{i_1}^{\nu},
\nn\\[.3em]
\{g pp \}^{\mu\nu\rho\sigma}_{i_1 i_2} &=&
 g^{\mu\nu}p_{i_1}^{\rho}p_{i_2}^{\sigma}
+g^{\mu\rho}p_{i_1}^{\sigma}p_{i_2}^{\nu}
+g^{\mu\sigma}p_{i_1}^{\nu}p_{i_2}^{\rho}
+g^{\nu\rho}p_{i_1}^{\sigma}p_{i_2}^{\mu}
+g^{\rho\sigma}p_{i_1}^{\nu}p_{i_2}^{\mu}
+g^{\sigma\nu}p_{i_1}^{\rho}p_{i_2}^{\mu}
.
\label{eq:covbrace}
\eea
For illustration, we explicitly display the tensor decomposition of $T^{3,\mu\nu\rho\sigma} = C^{\mu\nu\rho\sigma}$
\bea
C^{\mu\nu\rho\sigma} &=&\sum_{i_1,i_2,i_3,i_4=1}^{2}
p^{\mu}_{i_1} p^{\nu}_{i_2} p^{\rho}_{i_2} p^{\sigma}_{i_2}C_{i_1 i_2
i_3 i_4 }+ \sum_{i_1,i_2=1}^{2} \{g p p\}^{\mu\nu\rho\sigma}_{i_1,i_2} C_{00 i_1
i_2} + \{g g\}^{\mu\nu\rho\sigma}C_{0000}
\eea
Note that in this convention terms like
\begin{equation}
\nonumber
p^{\mu}_{1} p^{\nu}_{1} p^{\rho}_{1} p^{\sigma}_{2}C_{1112}
+  p^{\mu}_{1} p^{\nu}_{1} p^{\rho}_{2} p^{\sigma}_{1}C_{1121}
+p^{\mu}_{1} p^{\nu}_{2} p^{\rho}_{1} p^{\sigma}_{1} C_{1211}
+p^{\mu}_{2} p^{\nu}_{1} p^{\rho}_{1} p^{\sigma}_{1} C_{2111}
\end{equation}
occur separately though they could be put together due to the symmetry of the tensor coefficients in all indices.

\section{Feynman parametrization}
The investigation of the divergent behaviour is most easily done by means of
Feynman parametrization. From its most general form
\be\label{genfeynpara}
\frac{1}{A_1^{m_1}A_2^{m_2}\dots A_n^{m_n}} = 
\int_0^1 dx_1 dx_2 .. dx_n
\, \delta \bigg( 1 - \sum_{i=1}^n x_i \bigg)\, 
\frac{\prod x_i^{m_i-1}}{(\sum_i A_i x_i)^n} \frac{\Gamma(m_1+..m_n)}{\Gamma(m_1)..\Gamma(m_n)}
\ee
we obtain for the denominators of eq.\ref{def}
\bea\label{ourfeynpara}
\frac{1}{N_0\dots N_{N-1}}&=&\int_0^1 dx_0 dx_1 .. dx_{N-1}\,
\delta \bigg( 1 - \sum_{i=0}^{N-1} x_i \bigg)\,\frac{\Gamma(N)}{(\sum_{i=0}^{N-1} N_i
x_i)^N}\\
&=& \nonumber (N-1)! \int_0^1 dx_1 \int_0^{1-x_1} dx_2 \dots \int_0^{1-x_1-\dots-x_{N-2}}
dx_{N-1} \\
&& \times \bigg\{ N_0(1-\sum_{i=1}^{N-1} x_i) + \sum_{i=1}^{N-1} x_i
N_i \bigg\}^{-N}
\eea
By denoting the integration over the $N$-dimensinal simplex as $\int dS_{N}$
we can thus write the general one loop tensor N-point integral (\ref{def}) as
\be
T^{N, \mu_1 \dots \mu_{P}} =
\frac{(2\pi \mu)^{4-D}}{i \pi^2} (N-1)! \int d^D q \, q^{\mu_1}..q^{\mu_P} \int dS_{N-1} \bigg\{ N_0(1-\sum_{i=1}^{N-1} x_i) + \sum_{i=1}^{N-1}
x_i N_i \bigg\}^{-N}.
\ee
Explicitly inserting the $N_i$'s, completing the square
and performing a shift in the integration variable $q$ gives
\bea\nonumber
T^{N, \ \mu_1 \dots
\mu_{P}} &=& \frac{(2\pi \mu)^{4-D}}{i \pi^2} (N-1)! \int d^D q \, (q -
\sum_{i=1}^{N-1} x_i p_i)^{\mu_1}\dots (q - \sum_{i=1}^{N-1} x_i p_i)^{\mu_P}\\ && \int
dS_{N-1} \Bigg\{ q ^2 - A^{(N)}(x_i,(p_ip_j),m_i^2) +i\eta
\Bigg\}^{-N}\label{feyntensor}
\eea
with
\be
A^{(N)}(x_i,(p_ip_j),m_i^2) = 
\sum_{i=1}^{N-1}\left( p_i x_i \right)^2 +
\sum_{i=1}^{N-1}\sum_{j(>i)=1}^{N-1} 2 p_ip_j x_ix_j 
-\sum_{i=1}^{N-1} x_if_i
+ m_0^2
\ee
and
\be
f_i = p_i^2-m_i^2+m_0^2
\ee
as result that we want to use for further calculation of the divergent parts.

\section{Calculation of the UV-divergent parts}
Regularization of loop integrals can be done with the aid of various techniques. Dimensional regularization has become the standard scheme to deal with UV divergences \cite{'tHooft:1972fi},\cite{Bollini:1972bi},\cite{Ashmore:1972uj}.
One could in principle also handle IR divergences in the framework of dimensional regularization, but then both are parameterized by $\frac{1}{\varepsilon}$-terms and become indistinguishable. To separate them we assume the introduction of small regulator masses $\lambda_i$ which guarantee IR finiteness. Thus, we need not to worry about the infra-red behaviour in the following calculation.

At first, we investigate the UV-divergent part of a tensor coefficient proportional to metric tensors only. Later, we reduce the calculation for coefficients belonging to momenta  and metric tensors to the upper case. In order to extract the coefficient proportional to metric tensors only from the Feynman parametrization of a general one-loop tensor integral (with an even number of indices) one just drops all terms $\sum x_i p_i$ in eq.(\ref{feyntensor}), that is
\be\label{tensor00}
\{\underbrace{g..g}_n\}^{\mu_1 .. \mu_{2n}}\,
T^N_{\underbrace{\sst 0..0}_{2n}}=\frac{(2\pi \mu)^{4-D}}{i \pi^2} (N-1)!
\int d^D q \, q^{\mu_1}\dots q^{\mu_{2n}} \int
dS_{N-1} \Big\{ q ^2 - A^{(N)} +i\eta\Big\}^{-N }.
\ee
Contracting both sides of eq.(\ref{tensor00}) with $n$ metric tensors yields
\be\label{metricpref}
g_{\mu_1\mu_2}\dots g_{\mu_{2n-1}\mu_{2n}} \{g..g\}^{\mu_1..\mu_{2n}} = D(D+2)\dots \big(D+2(n-1)\big)
\ee
on the left hand side and generates $n$ times $q^2$ on the right hand side. Using
\be\label{prefac}
\prod_{i=1}^n \frac{1}{D + 2(i-1)}=\prod_{i=1}^n \frac{1}{2(i+1)-\varepsilon}=
\prod_{i=1}^n \frac{1}{2(i+1)} \big(1+ O(\varepsilon)\big)=
\frac{2^{-n}}{(n+1)!}\big(1+ O(\varepsilon)\big)
\ee
and
\begin{equation}\nn
(2\pi \mu)^{4-D}=(2\pi \mu)^{\varepsilon}= e^{\varepsilon \ln 2\pi\mu}
= 1 + \varepsilon \ln 2\pi\mu +\dots
\end{equation}
we obtain
\be
T^N_{\underbrace{\sst 0..0}_{2n}}=
\frac{2^{-n}}{(n+1)!}
\frac{\big(1+ O(\varepsilon)\big)}{i \pi^2} (N-1)!\int
dS_{N-1} \int d^D q \,
\left( q^2 \right)^n  \Big\{ q ^2 - A^{(N)} +i\eta\Big\}^{-N }
\ee
The next step consists of carrying out the $q$ integration. Therefore, we transform the integrand with the help of the binomial theorem (for better readability we leave out the upper index in $A^{(N)}$ until  eq.\ref{UVpart_metrictensorcoeff})
\bea
&&\left( q^2 \right)^n \big\{ q ^2 - A +i\eta\big\}^{-N} =
(q ^2-A+i\eta+A-i\eta)^n\,\big\{q ^2 - A +i\eta
\big\}^{-N} =\nonumber\\
&&\sum^n_{k=0}{n \choose k}\big(A-i\eta\big)^k
\bigg\{ q^2 -A+i\eta\bigg\}^{n-k}
\big\{ q ^2 - A +i\eta\big\}^{-N}
\eea
which yields for the tensor coefficient
\be
T^N_{\underbrace{\sst 0..0}_{2n}}=
\frac{2^{-n}}{(n+1)!}
\frac{\big(1+ O(\varepsilon)\big)}{i \pi^2} (N-1)!\int
dS_{N-1} \sum^n_{k=0}{n \choose k}\big(A-i\eta\big)^k\int d^D q \,
\Big\{ q ^2 - A +i\eta\Big\}^{-N-k+n }
\ee
To perform the $q$-integration we distinguish between two cases
\bea
&& \textrm{1.Case:} \qquad -N-k+n = \beta \ge 0 \nonumber
\eea
The pole prescription becomes unnecessary in this case and we use the rules for integration over a $D$-dimensional space, that can be found for example in \cite{collins}.
\bea
&& \int d^D q \,\Big\{ q ^2 - A \Big\}^{\beta}=
\int d^D q \,\sum_{\alpha=0}^{\beta} c_{\alpha}(A)\,(q ^2)^{\alpha}=
\sum_{\alpha=0}^{\beta} c_{\alpha}(A)\,\int d^D q \,(q^2)^{\alpha}=0 \nonumber
\eea
Explicitly, we exploited linearity and applied the result $\int d^D q \,(q^2)^{\alpha}=0 ,\forall \alpha$.
\bea
&& \textrm{2.Case:} \qquad -N-k+n= -\beta < 0 \nonumber
\eea
In this case, we deal with the standard integral $I_{\beta}(A)$ given by
\bea
&&
\int d^D q \,\frac{1}{(q ^2 - A + i\eta)^{\beta}}=I_{\beta}(A)=
i \pi^{\frac{D}{2}}(-1)^{\beta}\frac{\Gamma(\beta-\frac{D}{2})}{\Gamma(\beta)}
\big(A-i\eta\big)^{\frac{D}{2}-\beta}\nonumber
\eea
A detailed calculation of this integral is executed in most standard books on quantum field theory that use dimensional regularization, for example in \cite{boehm}.

After omitting the vanishing parts with the positive exponent our tensor coefficient reads
\be
T^N_{\underbrace{\sst 0..0}_{2n}}=
\frac{2^{-n}}{(n+1)!}
\frac{\big(1+ O(\varepsilon)\big)}{i \pi^2} (N-1)!\int
dS_{N-1} \sum^n_{k=n-N+1}{n \choose k}\big(A-i\eta\big)^k \,
I_{N+k-n}(A)
\ee
In dimensions close to 4, UV divergences only occur for $I_1$ and $I_2$, i.e for $k=n-N+1$ and $k=n-N+2$, the other $I_j$'s are UV-finite. Thus, we are left with
\bea\nonumber
T^N_{\underbrace{\sst 0..0}_{2n}}&=&
\frac{2^{-n}}{(n+1)!}
\frac{1}{i \pi^2} (N-1)!\int
dS_{N-1}
\Bigg[  {n \choose  n-N+1}
\big( A-i\eta \big)^{n-N+1}\,I_1(A) \\
&&
{n \choose n-N+2}\big(A-i\eta\big)^{n-N+2}\,I_2(A)\Bigg]
+ \textrm{UV-finite terms} \nonumber
\eea
After inserting the explicit forms of $I_1$ and $I_2$ and using $\pi^{\frac{\varepsilon}{2}}=1+O(\varepsilon)$ and $\Gamma(z+1)=z\Gamma(z)$ we arrive at
\bea
T^N_{\underbrace{\sst 0..0}_{2n}}
&=&
\frac{2^{-n}}{(n+1)!}\Bigg[ \frac{n!}{(n-N+1)!}\frac{1}{1-\frac{\varepsilon}{2}}
+\frac{n!(N-1)}{(n-N+2)!}\Bigg]\nonumber\\
&&\int
dS_{N-1}(A-i\eta)^{n-N+2-\frac{\varepsilon}{2}}\,\Gamma\left(\frac{\varepsilon}{2}\right)
+ \textrm{UV-finite terms}\nonumber
\eea
By ingnoring UV-finite terms, the prefactor can be further simplified yielding
\bea
T^N_{\underbrace{\sst 0..0}_{2n}}&=&
\frac{2^{-n}}{(n-N+2)!}\int
dS_{N-1} \,(A-i\eta)^{n-N+2-\frac{\varepsilon}{2}}\,
\Gamma\left(\frac{\varepsilon}{2}\right)
+ \textrm{UV-finite terms}
\eea
By taking into account that
\bea
(A-i\eta)^{-\frac{\varepsilon}{2}}=1- \frac{\varepsilon}{2}\ln
(A-i\eta)+\dots \nonumber
\eea
and using
\be
\Gamma\left(\frac{\varepsilon}{2}\right)=\frac{2}{\varepsilon}-\gamma_E +
O\left(\frac{\varepsilon}{2}\right),\quad \textrm{for } \varepsilon \rightarrow 0,
\quad \gamma_E= 0,5722.. \textrm{(Euler's constant)}
\ee
we get
\bea
T^N_{\underbrace{\sst 0..0}_{2n}}&=&
\frac{2^{-n}}{(n-N+2)!}\,\frac{2}{\varepsilon}\, \int
dS_{N-1} \,(A-i\eta)^{n-N+2}
+ \textrm{UV-finite terms}
\eea
and finally for the UV-divergent part (we can omit the $i\eta$ now, because for all one loop N-point tensor functions that are UV-divergent the relation $n-N+2\ge 0$ holds)
\bea
\label{UVpart_metrictensorcoeff}
\big( D-4 \big)\, T^N_{\underbrace{\sst 0..0}_{2n}}&=&
- \frac{2^{-(n-1)}}{(n-N+2)!}\int dS_{N-1} \,A^{(N)}(x_i, p_ip_j ,m_i^2)^{n-N+2}
\eea
With the help of this special result we can now investigate the divergence of a general tensor coefficient that we extract from equation
(\ref{feyntensor}).
\bea
&&\{\underbrace{g..g}_n\}^{\mu_1 .. \mu_{2n}}
\underbrace{p_1^{\mu_{2n+1}} .. p_1^{\mu_{2n+k_1}}}_{k_1}\,
\dots
\underbrace{p_{N-1}^{\mu_{2n+k_1+..+k_{N-2}+1}} .. p_{N-1}^{\mu_{2n+k_1+..+k_{N-1}}}}_{k_{N-1}}\,
T^N_{\underbrace{\sst 0..0}_{2n}
\underbrace{\sst 1..1}_{k_1}..
\underbrace{\sst N-1..N-1}_{k_{N-1}}}=\nonumber\\
&&
\frac{(2\pi \mu)^{4-D}}{i \pi^2} (N-1)!
\int d^D q \, \int dS_{N-1}\,
q^{\mu_1}\dots q^{\mu_{2n}}\,
\big(-x_1 p_1\big)^{\mu_{2n+1}}..\big(-x_1 p_1\big)^{\mu_{2n+k_1}}
\dots \nonumber\\
&&
\big(-x_{N-1} p_{N-1}\big)^{\mu_{2n+k_1+..+k_{N-2}+1}}..
\big(-x_{N-1} p_{N-1}\big)^{\mu_{2n+k_1+..+k_{N-1}}}
\Big\{ q ^2 - A^{(N)} +i\eta\Big\}^{-N }
\eea
The external momenta only determine the form of the polynomial in the $x_i$'s. By comparing both sides one gets
\bea
&&\{\underbrace{g..g}_n\}^{\mu_1 .. \mu_{2n}}\,
T^N_{\underbrace{\sst 0..0}_{2n}
\underbrace{\sst 1..1}_{k_1}..
\underbrace{\sst N-1..N-1}_{k_{N-1}}}=\nonumber\\
&&
\frac{(2\pi \mu)^{4-D}}{i \pi^2} (N-1)!
\, \int dS_{N-1}\,\left(\prod_{i=1}^{N-1} (-x_i)^{k_i}\right)
\int d^D q \, q^{\mu_1}\dots q^{\mu_{2n}}\,
\Big\{ q ^2 - A^{(N)} +i\eta\Big\}^{-N } \nonumber
\eea
The $q$-integration can be performed as above and gives for the general one loop tensor N-point coefficient 
\be\label{gendiv}
\big( D-4 \big)\, T^N_{\underbrace{\sst 0..0}_{2n}
\underbrace{\sst 1..1}_{k_1}..
\underbrace{\sst N-1..N-1}_{k_{N-1}}}=
- \frac{2^{-(n-1)}}{(n-N+2)!}\,(-1)^{\sum_{i=1}^{N-1}k_i}\int dS_{N-1}(A^{(N)})^{n-N+2}
\,\prod_{i=1}^{N-1} x_i^{k_i}
\ee
For the integration over the simplex $S_{N-1}$, we first simplify the notation by identifying each of the $\frac{N(N+1)}{2}$ summands of $A^{(N)}$ with a term $a_l$
\be
A^{(N)}(x_i,(p_ip_j),m_i^2) \equiv \sum_{l=1}^{\frac{N}{2}(N+1)} a_l = \underbrace{\sum_{i=1}^{N-1}\left( p_i x_i \right)^2}_{N-1\, {\rm terms}}
 +
\underbrace{\sum_{i=1}^{N-1}\sum_{j>i=1}^{N-1} 2 p_ip_j x_ix_j }_{\frac{(N-1)(N-2)}{2}\,{\rm terms}}
-\underbrace{\sum_{i=1}^{N-1} x_if_i}_{N-1 \, {\rm terms}}
+ m_0^2,
\ee
introducing an index vector $\vec s$ of length $\frac{N(N+1)}{2}$
\be
\label{indexvec_s}
\vec s =\{ s_1,s_2, ..., s_{\frac{N(N+1)}{2}}\},
\ee
and setting $n-N+2=m$. Applying the multinomial theorem yields
\be
\big( D-4 \big)\, T^N_{\underbrace{\sst 0..0}_{2n}
	\underbrace{\sst 1..1}_{k_1}..
	\underbrace{\sst N-1..N-1}_{k_{N-1}}}=
\frac{(-1)^{1+\sum_{i=1}^{N-1}k_i} }{2^{(n-1)}m!}\, 
\int dS_{N-1}
\sum_{|\vec s| =m} {m \choose \vec s}
\,\prod_{l=1}^{\frac{N}{2}(N+1)} a_l^{s_l}
\,\prod_{i=1}^{N-1} x_i^{k_i}
\ee
where the multinomial coefficient is given by
\be
\label{multinomial_def}
{m \choose \vec s} = \frac{m!}{s_1! \cdot s_2! \cdot \dots \cdot s_{\frac{N(N+1)}{2}}!}
\ee
and the summation runs over all combinations of $s_l$ that sum up to $m$ denoted as
\be
\label{modulus_s_def}
|\vec s|:= s_1+s_2+ \cdots + s_{\frac{N(N+1)}{2}}=m.
\ee
By defining a vector $\vec b$ consisting of the summands $a_l$ without the $x_i$, i.e
\bea
\nonumber
\vec b&=&
 \{\, p_1^2,\ldots\ldots\ldots\ldots\ldots\ldots\ldots,p_{N-1}^2,\\
\nonumber&&  \,\, 2p_1p_2,\ldots\ldots\ldots,2p_1p_{N-1},\\
\nonumber&&  \,\, 2p_2p_3,\ldots,2p_2p_{N-1},\\
\nonumber&&  \,\, \vdots \qquad \qquad \iddots \\
\nonumber&&  \,\, 2p_{N-2}p_{N-1},\\
\label{summands_noX}&&  \,\, -f_1,\ldots\ldots\ldots\ldots\ldots\ldots,-f_{N-1},\, m_0^2 \, \},
\eea
we rewrite
\be
\big( D-4 \big)\, T^N_{\underbrace{\sst 0..0}_{2n}
	\underbrace{\sst 1..1}_{k_1}..
	\underbrace{\sst N-1..N-1}_{k_{N-1}}}= 
\frac{(-1)^{1+\sum_{i=1}^{N-1}k_i} }{2^{(n-1)}m!}\, 
\sum_{|\vec s| =m} {m \choose \vec s}
\,\prod_{l=1}^{\frac{N}{2}(N+1)} b_l^{s_l}
\int dS_{N-1}
\,\prod_{i=1}^{N-1} x_i^{t_i}
\ee
where the collected exponents belonging to a certain $x_i$ read in general
\be
\label{general_exponents}
t_i= k_i + 2 s_i
+ \sum_{j=1}^{i-1} s_{\frac{j}{2}(2N-3-j)+i}
+ \sum_{j=1}^{N-1-i} s_{\frac{i}{2}(2N-1-i)+j}
+ s_{\frac{N(N-1)}{2}+i}.
\ee
We denote the sum over the $t_i$ as $u$
\be
\label{exponent_sum}
u:=\sum_{i=1}^{N-1} t_i= \sum_{i=1}^{N-1} k_i + 2  \sum_{i=1}^{\frac{N(N-1)}{2}} s_i
+  \sum_{i=\frac{N(N-1)}{2}+1}^{\frac{N(N+1)}{2}-1} s_i
\ee
and make use of 
\be
\int dS_N \prod_{i=1}^{N} x_i^{r_i}=\frac{\prod_{i=1}^{N} r_i!}{(N+\sum_{i=1}^{N}r_i)!}
\ee
to finally obtain for the UV-divergent part of a general one loop tensor coefficient
\be
\label{UVpart_general_res}
\big( D-4 \big)\, T^N_{\underbrace{\sst 0..0}_{2n}
	\underbrace{\sst 1..1}_{k_1}\dots
	\underbrace{\sst N-1..N-1}_{k_{N-1}}}= 
\frac{(-1)^{1+\sum_{i=1}^{N-1}k_i} }{2^{(n-1)}m!}\, 
\sum_{|\vec s| =m} {m \choose \vec s}
\, \prod_{l=1}^{\frac{N}{2}(N+1)} b_l^{s_l}
\, \frac{\prod_{i=1}^{N-1} t_i!}{(N-1+u)!}
\ee 
where $m=n-N+2$ and all necessary abbreviations and definitions are given by eqs.(\ref{indexvec_s}), (\ref{multinomial_def}), (\ref{modulus_s_def}), (\ref{summands_noX}), (\ref{general_exponents}), and (\ref{exponent_sum}).

\section{Explicit Formulae for $N=1,2,...,6$}
The result eq.(\ref{gendiv}) is very compact and can be applied to all one loop N-point tensor integrals, but for practical purpose the explicit formulae for the A-, B-, C-, ... functions can be useful as well.
\subsection{A-functions}
For 1-point tensor integrals $T_{...}^1 =A_{...}$ the expressions become very
simple since there are no external momenta. By recognizing 
\bea
\nonumber
k_i = 0 \, \forall i, \quad m=n+1, \quad \vec s ={s_1}, \quad \vec b ={m_0^2}, \quad t_i = 0 \, \forall i, \quad u=0
\eea
we get from eq.\ref{UVpart_general_res}
\be
\big( D-4 \big) A_{\underbrace{\sst 0..0}_{2n}}= 
\frac{-1}{2^{(n-1)}(n+1)!}\, 
\sum_{s_1 =n+1} {n+1 \choose s_1}
\, (m_0^2)^{s_1}= -\frac{ m_0^{2n+2}}{2^{(n-1)}(n+1)!}
\ee 
Note that the UV-divergent part of the scalar integral $A_0$ is obtained by setting $n=0$.

\subsection{B-functions}
For 2-point tensor coefficients $T_{...}^2=B_{...}$ we have
\bea
\nonumber
m=n, \quad  \vec s =\{s_1,s_2,s_3\}, \quad \vec b=\{p_1^2, -f_1, m_0^2\}, \quad
t_1 = k_1 +2 s_1 +s_2, \quad u=t_1
\eea
yielding
\be
\label{general_B}
\big( D-4 \big) B_{\underbrace{\sst 0..0}_{2n}
\underbrace{\sst 1..1}_{k_1}}= 
\frac{(-1)^{k_1+1} }{2^{(n-1)} n!}\, 
\sum_{s_1+s_2+s_3 =n} {n \choose s_1,s_2,s_3}
\,   
\frac{(p_1^2)^{s_1}  \,  (-f_1)^{s_2} \,  (m_0^2)^{s_3}}{k_1 +2 s_1 +s_2 +1}
\ee
This expression can be also found in \cite{PackageX}. Since 2-point tensor functions are usually denoted with arguments $B_{...}(p_1^2,m_1^2,m_0^2)$ we reinsert $f_1=p_1^2-m_1^2+m_0^2$ and get 
\be
\label{general_B_nofi}
\big( D-4 \big) B_{\underbrace{\sst 0..0}_{2n}
	\underbrace{\sst 1..1}_{k_1}}= 
\frac{(-1)^{k_1+1} }{2^{(n-1)} n!}\, 
\sum_{v_1+ ..+v_5 =n} {n \choose v_1, .., v_5}
\,   (-1)^{v_2+v_4}
\frac{(p_1^2)^{v_1+v_2}  \,  (m_1^2)^{v_3}   \,  (m_0^2)^{v_4+v_5}}{k_1 +2 v_1 +v_2+v_3+v_4 +1}
\ee
To avoid the combinatorics required for the summation in the multinomial theorem, it is possible to alternatively write eq.\ref{general_B} with binomial coefficients
\be
\label{general_B_binomial}
\big( D-4 \big) B_{\underbrace{\sst 0..0}_{2n}
	\underbrace{\sst 1..1}_{k_1}}= 
\frac{(-1)^{k_1+1} }{2^{(n-1)} n!}\, 
\sum_{l_1=0}^n \sum_{l_2=0}^{l_1} {n \choose l_1} {l_1 \choose l_2}
\,   
\frac{(p_1^2)^{l_2}  \,  (-f_1)^{l_1-l_2} \,  (m_0^2)^{n-l_1}}{k_1 +l_1 +l_2 +1}
\ee
In the appendix, UV-divergent parts of B-functions up to rank 8 are listed explicitly.

\subsection{C-functions}

For 3-point tensor coefficients $T_{...}^3=C_{...}$ we have $\frac{N}{2}(N+1)=6$ and the quantities used in eq.\ref{UVpart_general_res} become
\bea
&&m=n-1, \quad  \vec s =\{s_1,s_2,s_3,s_4,s_5,s_6\}, \quad \vec b=\{p_1^2,\, p_2^2,\, 2p_1p_2, -f_1, -f_2, m_0^2\}, \\
&&t_1 = k_1 +2 s_1 +s_3+s_4, \quad t_2 = k_2 +2 s_2 +s_3 + s_5, \quad u=t_1 +t_2
\eea
Thus, the UV-divergent part reads
\bea
\nonumber
&&\left(D-4 \right)C_{\underbrace{\sst 0..0}_{2n}
\underbrace{\sst 1..1}_{k_1}
\underbrace{\sst 2..2}_{k_2}}=
\frac{(-1)^{k_1+k_2+1} }{2^{(n-1)} (n-1)!}\, 
\sum_{s_1+..+s_6 = n-1} {n-1 \choose s_1,..,s_6} (p_1^2)^{s_1}  (p_2^2)^{s_2} (2p_1p_2)^{s_3} \cdot\\
\label{general_C}
&&
\qquad\qquad\qquad\qquad (-f_1)^{s_4} (-f_2)^{s_5} \, (m_0^2)^{s_6} \,
\frac{(k_1 +2 s_1 +s_3+s_4)!(k_2 +2 s_2 +s_3 + s_5)!}{(k_1+k_2 +2(s_1 +s_2 + s_3) +s_4+s_5 +2)!}
\eea
which can be also found in \cite{PackageX}. If the $f_i=p_i^2-m_i^2+m_0^2$ are reinserted and $2p_1p_2$ is written as $p_1^2+p_2^2-p_{12}$, where $p_{12}=(p_1-p_2)^2$, we obtain with $\vec v =\{v_1,v_2,\dots, v_{12}\}$ and $|\vec v| := \sum v_i$
\bea
\nonumber
&&\left(D-4 \right)C_{\underbrace{\sst 0..0}_{2n}
	\underbrace{\sst 1..1}_{k_1}
	\underbrace{\sst 2..2}_{k_2}}=
\frac{(-1)^{k_1+k_2+1} }{2^{(n-1)} (n-1)!}\, 
\sum_{|\vec v| = n-1} {n-1 \choose \vec v} (-1)^{v_5+v_6+v_8+v_9+v_{11}} \cdot \qquad\qquad\qquad\\
\nonumber
&&
\qquad\qquad\qquad\qquad \qquad\quad \, \, 
(p_1^2)^{v_1+v_3+v_6}  (p_2^2)^{v_2+v_4+v_9} (p_{12})^{v_5} 
(m_1^2)^{v_7} (m_2^2)^{v_{10}} \, (m_0^2)^{v_8+v_{11}+v_{12}}\cdot \\
\label{general_C_nofi}
&&
\qquad \frac{(k_1 +2 v_1 +v_3+v_4+v_5+v_6+v_7+v_8)!(k_2 +2 v_2 +v_3+v_4+v_5 + v_9+v_{10}+v_{11})!}{(k_1+k_2 +2 \sum_{i=1}^{5} v_i + \sum_{i=6}^{11} v_i +2)!}
\eea
An alternative expression for eq.\ref{general_C} using binomial coefficients is given by
\bea
\nonumber
\left( D-4 \right) C_{\underbrace{\sst 0..0}_{2n}
\underbrace{\sst 1..1}_{k_1}\underbrace{\sst 2..2}_{k_2}} &=& 
\frac{(-1)^{k_1+k_2+1} }{2^{(n-1)} (n-1)!} \,
\sum_{l_1=0}^{n-1} \sum_{l_2=0}^{l_1} \sum_{l_3=0}^{l_2}\sum_{l_4=0}^{l_3}\sum_{l_5=0}^{l_4} 
{n-1 \choose l_1} {l_1 \choose l_2}  {l_2 \choose l_3}  
{l_3 \choose l_4} \cdot\\ \nonumber
&&
  {l_4 \choose l_5} (p_1^2)^{l_5} (p_2^2)^{l_4-l_5}
(2 p_1p_2)^{l_3-l_4} 
(-f_1)^{l_2-l_3} (-f_2)^{l_1-l_2} 
(m_0^2)^{n-1-l_1}\cdot \\
&&
\frac{(k_1 + 2l_5-l_4+l_2)!(k_2 -2l_5 + l_4 + l_3 -l_2 + l_1)!}{(k_1 + k_2 + l_1 + l_3 +2)!}
\label{general_C_binomial}
\eea 
In the appendix, UV-divergent parts of C-functions up to rank 8 are shown explicitly.

\subsection{D-functions}
In case of four point tensor coefficients $T^4_{...}=D_{...}$ the length of the index vector $\vec s$ and of $\vec b$ amounts to $\frac{N}{2}(N+1)=10$ and the quantities used in eq.\ref{UVpart_general_res} are given by
\bea
&& m=n-2, \quad  \vec s =\{s_1,s_2,s_3,s_4,s_5,s_6,s_7,s_8,s_9,s_{10}\}\\
&& \vec b=\{p_1^2,\, p_2^2,\, p_3^2,\, 2p_1p_2, 2p_1p_3, 2p_2p_3, -f_1, -f_2,-f_3, m_0^2\}, \\
&& t_1 = k_1 +2 s_1 +s_4+s_5+s_7, \quad t_2 = k_2 +2 s_2 +s_4 + s_6 +s_8, \\
&& t_3 = k_3 +2 s_3 +s_5 + s_6 +s_9, \quad u=t_1 +t_2+t_3
\eea
yielding 
\bea
\nonumber
&&\left(D-4 \right)D_{\underbrace{\sst 0..0}_{2n}
	\underbrace{\sst 1..1}_{k_1}
	\underbrace{\sst 2..2}_{k_2}\underbrace{\sst 3..3}_{k_3}}=
\frac{(-1)^{k_1+k_2+k_3+1} }{2^{(n-1)} (n-2)!}\, 
\sum_{s_1+..+s_{10} = n-2} {n-2 \choose s_1,..,s_{10}}  \cdot \\
\nonumber && \qquad\quad
(p_1^2)^{s_1} \, (p_2^2)^{s_2} \, (p_3^2)^{s_3}
(2p_1p_2)^{s_4}(2p_1p_3)^{s_5}(2p_2p_3)^{s_6} 
(-f_1)^{s_7} (-f_2)^{s_8} (-f_3)^{s_9} \, (m_0^2)^{s_{10}}  \cdot\\
\label{general_D}
&& \qquad\quad
\frac{(k_1 +2 s_1 +s_4+s_5+s_7)!(k_2 +2 s_2 +s_4 + s_6 +s_8)!(k_3 +2 s_3 +s_5 + s_6 +s_9)!}{(k_1+k_2+k_3 +2(s_1 +s_2 + s_3 +s_4+s_5 +s_6) +s_7+ s_8 + s_9+3)!}
\eea

This expression can be rewritten using binomial coefficients as
\bea
\nonumber
&&\left(D-4 \right)D_{\underbrace{\sst 0..0}_{2n}
	\underbrace{\sst 1..1}_{k_1}
	\underbrace{\sst 2..2}_{k_2}\underbrace{\sst 3..3}_{k_3}}=
\frac{(-1)^{k_1+k_2+k_3+1} }{2^{(n-1)} (n-2)!}\, 
\sum_{l_1=0}^{n-2} \sum_{l_2=0}^{l_1} \sum_{l_3=0}^{l_2}\sum_{l_4=0}^{l_3}\sum_{l_5=0}^{l_4} \sum_{l_6=0}^{l_5} \sum_{l_7=0}^{l_6} \sum_{l_8=0}^{l_7} \sum_{l_9=0}^{l_8} \\
 \nonumber &&
 \qquad\qquad
{n-2 \choose l_1} {l_1 \choose l_2}  {l_2 \choose l_3}  
{l_3 \choose l_4}{l_4 \choose l_5}{l_5 \choose l_6}{l_6 \choose l_7}{l_7 \choose l_8}{l_8 \choose l_9} 
(p_1^2)^{l_9}(p_2^2)^{l_8-l_9}(p_3^2)^{l_7-l_8}\cdot
\\ \nonumber
&&
 \qquad\qquad
(2 p_1p_2)^{l_6-l_7} (2 p_1p_3)^{l_5-l_6} (2 p_2p_3)^{l_4-l_5} 
(-f_1)^{l_3-l_4} (-f_2)^{l_2-l_3} (-f_3)^{l_1-l_2} 
(m_0^2)^{n-2-l_1}\cdot\\
\nonumber
&& \qquad\qquad
\frac{(k_1 + 2 l_9 - l_7 + l_5 - l_4 + l_3)!
	(k_2 - 2 (l_9 - l_8)  - l_7 + l_6 - l_5 + l_4 - l_3 + l_2)! 
}{(k_1 + k_2 + k_3 + l_4 + l_1 +3)!}\cdot \\
&&  \qquad\qquad	(k_3 -2 (l_8 -l_7) - l_6 + l_4 - l_2+ l_1)!
\label{general_D_binomial}
\eea 
In the appendix, UV-divergent parts of D-functions up to rank 8 are listed explicitly.

\subsection{E- and F- functions}

In a four dimensional space-time, 6-point integrals  $T^6_{...}=F_{...}$ can be reduced to 5-point integrals $T^5_{...}=E_{...}$ that can be further reduced to 4-point functions \cite{Denner:2005nn}, \cite{Denner:2002ii}, \cite{Me65}.
Nevertheless, we want to provide the explicit form of the quantities required in the generic expression eq.\ref{UVpart_general_res} in order to demonstrate
the general applicability of our result. Additionally, UV-divergent parts of high ranked coefficients can occur in reduction procedures for $E$- and $F$-functions \cite{Denner:2005nn}.
 
For $E$-functions, the length of the index vector $\vec s$ and of $\vec b$ amounts to $\frac{N}{2}(N+1)=15$ and we get for the abbreviation used in eq.(\ref{UVpart_general_res})
\bea
\nonumber
&& m=n-3, \quad  \vec s =\{s_1,s_2,s_3,s_4,s_5,s_6,s_7,s_8,s_9,s_{10},s_{11},s_{12},s_{13},s_{14},s_{15}\},
\\
\nonumber&&
 \vec b=\{p_1^2,\, p_2^2,\, p_3^2,\, p_4^2,\,2p_1p_2, 2p_1p_3, 2p_1p_4, 2p_2p_3,2p_2p_4,2p_3p_4, -f_1, -f_2,-f_3,-f_4, m_0^2\}, \\
 \nonumber&&
 t_1 = k_1 +2 s_1 +s_5+s_6+s_7+s_{11}, \quad t_2 = k_2 +2 s_2 +s_5 + s_8 +s_9 +s_{12}, \\
 \nonumber&&
t_3 = k_3 +2 s_3+ s_6 +s_8 + s_{10} +s_{13}, \quad t_4 = k_4 +2 s_4 +s_7 + s_9 +s_{10}+s_{14}\\
&& u =t_1+t_2+t_3+t_4
\eea
 For $F$-functions, the length of the index vector $\vec s$ and of $\vec b$ amounts to $\frac{N}{2}(N+1)=21$ and we obtain
 \bea
 \nonumber
\vec s &=& \{s_1,s_2,s_3,s_4,s_5,s_6,s_7,s_8,s_9,s_{10},s_{11},s_{12},s_{13},s_{14},s_{15},s_{16},s_{17},s_{18},s_{19},s_{20},s_{21}\},
\\
\nonumber
\vec b &=& \{p_1^2,\, p_2^2,\, p_3^2,\, p_4^2,\, p_5^2,\,2p_1p_2, 2p_1p_3, 2p_1p_4, 2p_1p_5, 2p_2p_3,2p_2p_4, 2p_2p_5,2p_3p_4,2p_3p_5, \\
\nonumber && 2p_4p_5, -f_1, -f_2,-f_3,-f_4,-f_5, m_0^2\}, \quad m=n-4 \\
 \nonumber
 t_1&=& k_1 +2 s_1 +s_6+s_7 + s_8+s_9+s_{16}, \qquad 
 t_2 = k_2 +2 s_2 + s_6 +s_{10}+s_{11}  +s_{12}+s_{17}, \\
 \nonumber
 t_3&=& k_3 +2 s_3+ s_7 + s_{10} +s_{13}+s_{14}+s_{18}, \quad 
 t_4 =  k_4 +2 s_4 +s_8 +s_{11} + s_{13} + s_{15}+ s_{19}, \\
 t_5&=& k_5 +2 s_5 +s_9 +s_{12} + s_{14} + s_{15}+ s_{20}, \quad u =t_1+t_2+t_3+t_4+t_5
 \eea

\section{Conclusion}

In this work, the UV-divergent term of an arbitrary one-loop tensor coefficient in dimensional regularization is evaluated by using Feynman parametrization. 
The resulting formula is very compact and can be easily implemented. The computational effort is mainly determined by finding all possible index combinations over which the summation in the multinomial theorem runs. 

Compared to recursive schemes existing for the calculation of these UV-divergent parts, our result can be implemented with less programming effort and benefits from the obvious advantages of a closed expression over a recursion formula, i.e no start values needed or direct access to coefficient functions of high rank.

Such a generally valid result contributes to the completion of investigations on UV-divergent parts of one-loop tensor coefficients. For practical purposes, the formula proves its usefulness in reduction procedures or for approximate formulas of tensor coefficients whose required accuracy is not known a priori. 

In the appendix, we provide the most extensive tables currently available in literature allowing to straight forwardly look up UV-divergent parts of tensor coefficients of B-, C-, D-, E-functions up to rank 8 and of F-functions up to rank 10.
\newline
\newline

\vspace{10mm}
\huge
\textbf{Appendix}
\normalsize
\newline

\begin{appendix}
	
Here, we want to give the explicit expressions for the UV-divergent parts of B-, C-, D-, and E-functions up rank 8 and for F-functions up the rank 10 thus representing the most extensive collection of UV-divergent parts currently available in literature. 

The abbreviation $p_{ij}$ always stands for $(p_i-p_j)^2$. 

\section{UV-divergent parts of B-functions up to rank 8}

The explicit formulae for 2-point tensor coefficients up to rank 8 read ($B_0$ is obtained from eq.\ref{general_B} for $n=0$)

\begin{eqnarray*}
	(D-4)B_{0}&=&-2,\quad
	(D-4)B_{1}=1\\
	(D-4)B_{00}&=&\frac{1}{6}\left(-3 m_0^2 - 3 m_1^2 + p_1^2\right),\quad
	(D-4)B_{11}= -\frac{2}{3} \\
	(D-4)B_{001}&=&\frac{2 m_0^2 + 4 m_1^2 - p_1^2}{12},\quad
	(D-4)B_{111}=\frac{1}{2}\\
	(D-4)B_{0000}&=&\frac{-10 m_0^4 - 10 m_1^4 + 5 m_1^2 p_1^2 -
		p_1^4 + 5 m_0^2 \left( -2 m_1^2 + p_1^2 \right) }{120}\\
	(D-4)B_{0011}&=&\frac{-5 m_0^2 + 3 \left( -5 m_1^2 + p_1^2 \right) }{60},\quad
	(D-4)B_{1111}=- \frac{2}{5}  \\
	(D-4)B_{00001}&=&\frac{5 m_0^4 + 15 m_1^4 - 6 m_1^2\ p_1^2 + 
		p_1^4 + 2 m_0^2 \left( 5 m_1^2 - 2 p_1^2 \right) }{240}\\
	(D-4)B_{00111}&=&\frac{3 m_0^2 + 12 m_1^2 - 2 p_1^2}{60},\quad
	(D-4)B_{11111}=\frac{1}{3}\\
	(D-4)B_{000000}&=&\frac{-35 m_0^6 - 35 m_1^6 + 
		21 m_1^4 p_1^2 - 7 m_1^2 p_1^4 + p_1^6 }{3360} +\\
	&&\frac{ - 7 m_0^4 \left( 5 m_1^2 - 3 p_1^2 \right)-
		7 m_0^2 \left( 5 m_1^4 - 4 m_1^2 p_1^2
		+ p_1^4 \right) }{3360}\\
	(D-4)B_{000011}&=&\frac{-7 m_0^4 + 7 m_0^2\ \left( -3 m_1^2 + 
		p_1^2 \right)  - 2 \left( 21 m_1^4
		- 7 m_1^2 p_1^2 + p_1^4 \right)  }{840}\\
	(D-4)B_{001111}&=&\frac{-7 m_0^2 + 5 \left( -7 m_1^2 + p_1^2 \right) }{210},\quad
	(D-4)B_{111111} = - \frac{2}{7}  \\
	(D-4)B_{0000001}&=&\frac{14 m_0^6 + 56 m_1^6 - 
		28 m_1^4 p_1^2 + 8 m_1^2 p_1^4 - p_1^6}{6720}+\\
	&&\frac{ 14 m_0^4 \left( 2 m_1^2 - p_1^2 \right) + m_0^2
		\left( 42 m_1^4 - 28 m_1^2 p_1^2 + 6 p_1^4  \right) }{6720}\\
	(D-4)B_{0000111}&=&\frac{14 m_0^4 + 8 m_0^2 \left( 7 m_1^2 -
		2 p_1^2 \right)  + 5 \left( 28 m_1^4 - 8 m_1^2
		p_1^2 + p_1^4 \right) }{3360} \\
	(D-4)B_{0011111}&=& \frac{4 m_0^2 + 24 m_1^2 - 3 p_1^2}{168} , \quad
	(D-4)B_{1111111} = \frac{1}{4}\\
	(D-4)B_{00000000}&=&\frac{-126 m_0^8 - 126 m_1^8 +
		84 m_1^6 p_1^2 - 36 m_1^4 p_1^4 +
		9 m_1^2 p_1^6 - p_1^8}{120960}+\\
	&&\frac{- 42 m_0^6 \left( 3 m_1^2
		- 2 p_1^2 \right) - 18 m_0^4 \left( 7 m_1^4
		- 7 m_1^2 p_1^2 + 2 p_1^4 \right)}{120960}+\\
	&&\frac{- 9 m_0^2 \left( 14 m_1^6 - 14 m_1^4 p_1^2
		+  6 m_1^2 p_1^4 - p_1^6 \right) }{120960}\\
	(D-4)B_{00000011}&=&\frac{-42 m_0^6 - 18 m_0^4 \left( 7 m_1^2 -
		3 p_1^2 \right) - 9 m_0^2 \left( 28 m_1^4 - 16 m_1^2\ p_1^2
		\right. }{60480}+ \\
	&& \frac{\left. +  3 p_1^4 \right)+ 5 \left( -84 m_1^6
		+ 36 m_1^4\ p_1^2 -
		9 m_1^2 p_1^4 + p_1^6 \right) }{60480}\\
	(D-4)B_{00001111}&=&\frac{-12 m_0^4 + 15 m_0^2 \left( -4 m_1^2 +
		p_1^2 \right) -5 \left( 36 m_1^4 - 9 m_1^2 p_1^2 +
		p_1^4 \right) }{5040}\\
	(D-4)B_{00111111}&=&\frac{-9 m_0^2 + 7 \left( -9 m_1^2 + p_1^2 \right)
	}{504},\quad
	(D-4)B_{11111111} =  -\frac{2}{9}
\end{eqnarray*} 

\section{UV-divergent parts of C-functions up to rank 8}

Tensor coefficients with $n=0$ are UV-finite, $p_{12}$ denotes $(p_1-p_2)^2$.

\begin{eqnarray*}
	(D-4) C_{00}&=&- \frac{1}{2},\quad
	(D-4) C_{00i}=\frac{1}{6} \\
	(D-4) C_{0000}&=&\frac{-4\ m_0^2 - 4\ m_1^2 - 4\ {{m_2}}^2 + p_1^2 + \
		{p_{12}} + {{p_2}}^2}{48}\\
	(D-4) C_{00ii}&=&- \frac{1}{12},\quad
	(D-4) C_{00ij} =-\frac{1}{24} \\
	(D-4) C_{0000i}&=&\frac{1}{240}\left[ 5 m_0^2 - 2 {p_{12}}
	+ \sum_{n=1}^2 \left( 5 m_n^2 - p_n^2 \right)\left(1+\delta_{in}\right)\right]\\
	(D-4) C_{00iii}&=&\frac{1}{20},\quad
	(D-4) C_{00iij} = \frac{1}{60}\\
	(D-4) C_{000000}&=&\frac{-15 m_0^4 - 15 m_1^4 - 15 {{m_2}}^4 +
		3 {{m_2}}^2\ p_1^2 - p_1^4 +
		6 {{m_2}}^2\ {p_{12}} - p_1^2\ {p_{12}} }{1440}+\\
	&&\frac{
		- {{p_{12}}}^2 + 6 {{m_2}}^2\ {{p_2}}^2 -
		p_1^2\ {{p_2}}^2 - {p_{12}}\ {{p_2}}^2 - {{p_2}}^4 +
		3 m_1^2 \left( -5 {{m_2}}^2 + 2 p_1^2 \right.}{1440}+\\
	&&\frac{\left. 2 {p_{12}} + 
		{{p_2}}^2 \right) +
		3 m_0^2\ \left( -5 m_1^2 - 5 {{m_2}}^2 + 2\ p_1^2 +
		{p_{12}} + 2 {{p_2}}^2 \right) }{1440}\\
	(D-4) C_{0000ii}&=&-\frac{1}{720}\left[ 6m_0^2 - 3 {p_{12}}
	+ \sum_{n=1}^2 \left( 6 m_n^2 - p_n^2 \right)\left(1+2\delta_{in}\right)\right]\\
	(D-4) C_{0000ij}&=&-\frac{1}{720}\left[ 3m_0^2 +2 {p_{12}}
	+ \sum_{n=1}^2 \left( 6 m_n^2 - p_n^2 \right)\right]\\
	(D-4) C_{00iiii}&=&- \frac{1}{30},\quad
	(D-4) C_{00iiij} = - \frac{1}{120},\quad
	(D-4) C_{00iijj} = - \frac{1}{180} \\
	(D-4) C_{000000i}&=&-\frac{1}{10080}\left[ 3 {p_{12}}^2 + 21 m_0^4 - 7{p_{12}}m_0^2
	+ {p_{12}} \sum_{n=1}^2 \left(p_n^2 - 7 m_n^2
	\right)\left(2+\delta_{in}\right)\right. \\
	&&\left. + 7m_0^2 \sum_{n=1}^2 \left( 3 m_n^2
	- p_n^2 \right)\left(1+\delta_{in}\right)\right. \\
	&&\left.
	+ \sum_{n,m=1}^2 \left( p_n^2 p_m^2 - 7 p_n^2 m_m^2 + 21 m_n^2 m_m^2 \right)
	\left(1+2\delta_{in}\right)\right]\\
	(D-4) C_{0000iii}&=&\frac{1}{1680}\left[ 7 m_0^2 - 4 {p_{12}}
	+ \sum_{n=1}^2 \left( 7 m_n^2 - p_n^2 \right)\left(1+3\delta_{in}\right)\right]\\
	(D-4) C_{0000iij}&=&\frac{1}{5040}\left[ 7 m_0^2 - 6 {p_{12}}
	+ \sum_{n=1}^2 \left( 7 m_n^2 - p_n^2 \right)\left(2+\delta_{in}\right)\right]\\
	(D-4) C_{00iiiii}&=&\frac{1}{42},\quad
	(D-4) C_{00iiiij} = \frac{1}{210},\quad
	(D-4) C_{00iiijj} = \frac{1}{420}\\
	(D-4) C_{00000000}&=&\frac{-168 m_0^6 - 168 m_1^6 - 168 {{m_2}}^6 + 28\
		{{m_2}}^4\ p_1^2 - 8 {{m_2}}^2\ p_1^4 + 3 p_1^6 }{161280}+ \\
	&& \frac{ 84 {{m_2}}^4\ {p_{12}} -
		16 {{m_2}}^2\ p_1^2\ {p_{12}} + 3 p_1^4\ {p_{12}} -
		24 {{m_2}}^2\ {{p_{12}}}^2 +
		3 p_1^2\ {{p_{12}}}^2  }{161280}+ \\
	&& \frac{ 3 {{p_{12}}}^3 + 84 {{m_2}}^4\ {{p_2}}^2 -
		16 {{m_2}}^2\ p_1^2\ {{p_2}}^2 + 3 p_1^4\ {{p_2}}^2 -
		24 {{m_2}}^2\ {p_{12}}\ {{p_2}}^2  }{161280}+ \\
	&& \frac{ 4 p_1^2\ {p_{12}}\ {{p_2}}^2 + 3 {{p_{12}}}^2\ {{p_2}}^2 -
		24 {{m_2}}^2\ {{p_2}}^4 +
		3 p_1^2\ {{p_2}}^4 + 3 {p_{12}}\ {{p_2}}^4 + 3 {{p_2}}^6 }{161280}+ \\
	&& \frac{ -
		28 m_0^4 \left( 6 m_1^2 + 6 {{m_2}}^2 - 3 p_1^2 -
		{p_{12}} - 3 {{p_2}}^2 \right)- 28 m_1^4 \left( 6 {{m_2}}^2 - 3 p_1^2
		\right. }{161280}+ \\
	&& \frac{\left. - 3 {p_{12}} - {{p_2}}^2 \right) -
		8 m_0^2 \left( 21 m_1^4 + 21 {{m_2}}^4 + 3 p_1^4 +
		2 p_1^2\ {p_{12}} +
		{{p_{12}}}^2   \right.}{161280}+ \\
	&& \frac{\left.+ 3 p_1^2\ {{p_2}}^2 + 2 {p_{12}}\ {{p_2}}^2 +
		3 {{p_2}}^4 +
		7 m_1^2 \left( 3 {{m_2}}^2 - 2 p_1^2 - {p_{12}} -
		{{p_2}}^2 \right)  \right.}{161280}+ \\
	&& \frac{\left.   -
		7 {{m_2}}^2  \left( p_1^2 + {p_{12}} + 2 {{p_2}}^2 \right)
		\right)  - 8 m_1^2\ \left( 21 {{m_2}}^4 + 3 p_1^4
		+ 3 {{p_{12}}}^2  \right.}{161280}+ \\
	&& \frac{\left.  2 {p_{12}}\ {{p_2}}^2 +
		{{p_2}}^4 - 7 {{m_2}}^2\ \left( p_1^2 + 2 {p_{12}} + {{p_2}}^2 \right)  +
		p_1^2\ \left( 3 {p_{12}} + 2 {{p_2}}^2 \right) \right) }{161280}\\
	(D-4) C_{000000ii}&=&\frac{1}{40320}\left[ -6 {p_{12}}^2 + 12{p_{12}}m_0^2
	- {p_{12}} \sum_{n=1}^2 \left(3 p_n^2 - 24 m_n^2
	\right)\left(1+\delta_{in}\right)\right. \\
	&&\left. -28 m_0^4 -4 m_0^2 \sum_{n=1}^2 \left( 7 m_n^2
	- 2 p_n^2 \right)\left(1+2\delta_{in}\right)\right. \\
	&&\left.
	-\frac{1}{2} \sum_{n,m=1}^2 \left( p_n^2 p_m^2 - 8 p_n^2 m_m^2 + 28 m_n^2 m_m^2 \right)
	\left(2 + \delta_{in}+ \delta_{im}+ 8 \delta_{in}\delta_{im}\right)\right]\\
	(D-4) C_{00000012}&=&\frac{-28\ m_0^4 - 84\ m_1^4 - 84\ {{m_2}}^4 +
		\ 16\ {{m_2}}^2\ p_1^2 - 3\ p_1^4 +
		48\ {{m_2}}^2\ {p_{12}}}{80640} + \\
	&&\frac{ - 6\ p_1^2\ {p_{12}} - 9\ {{p_{12}}}^2 + \ 24\ {{m_2}}^2\ {{p_2}}^2 -
		4\ p_1^2\ {{p_2}}^2 - 6\ {p_{12}}\ {{p_2}}^2 - 3\ {{p_2}}^4  }{80640}+ \\
	&&\frac{ -
		8\ m_0^2\ \left( 7\ m_1^2 + 7\ {{m_2}}^2 -
		2\ \left( p_1^2 + {p_{12}} + {{p_2}}^2 \right)  \right)   }{80640}+ \\
	&&\frac{ - 8\ m_1^2\ \left( 14\ {{m_2}}^2 - 3\ p_1^2 - 2\ \left( 3\ {p_{12}} \
		+ {{p_2}}^2 \right) \right) }{80640}\\
	(D-4) C_{0000iiii}&=&-\frac{1}{3360}\left[ 8 m_0^2 - 5 {p_{12}}
	+ \sum_{n=1}^2 \left( 8 m_n^2 - p_n^2 \right)\left(1+4\delta_{in}\right)\right]\\
	(D-4) C_{0000iiij}&=&-\frac{1}{6720}\left[ 4 m_0^2 - 4 {p_{12}}
	+ \sum_{n=1}^2 \left( 8 m_n^2 - p_n^2 \right)\left(1+\delta_{in}\right)\right]\\
	(D-4) C_{0000iijj}&=&-\frac{1}{20160}\left[ 8 m_0^2 - 9 {p_{12}}
	+ 3 \sum_{n=1}^2 \left( 8 m_n^2 - p_n^2 \right)\right]\\
	(D-4) C_{00iiiiii}&=& - \frac{1}{56}  ,\quad
	(D-4) C_{00iiiiij} =  - \frac{1}{336} ,\quad
	(D-4) C_{00iiiijj} =  - \frac{1}{840} \\
	(D-4) C_{00iiijjj}&=&  - \frac{1}{1120} \\
\end{eqnarray*}

\section{UV-divergent parts of D-functions up to rank 8}

Coefficients with $n < 2$ are UV-finite, $p_{ij}$ stands for $(p_i-p_j)^2$. 

\begin{eqnarray*}
(D-4)D_{0000}&=&- \frac{1}{12},\quad
(D-4)D_{0000i} = \frac{1}{48}\\
(D-4)D_{000000}&=&\frac{-5 m_0^2 - 5 m_1^2 - 5 {{m_2}}^2 -
            5 {{m_3}}^2 + p_1^2 + {p_{12}} + {p_{13}} +
            {{p_2}}^2 + {p_{23}} + {{p_3}}^2}{480}\\
(D-4)D_{0000ii}&=&- \frac{1}{120},\quad
(D-4)D_{0000ij} = - \frac{1}{240} \\
(D-4)D_{000000i}&=&\frac{1}{2880}\left[ 6 \sum_{n=0}^3 m_n^2 \left(1+\delta_{in}\right)
                  - \sum_{n=1}^3 p_n^2 \left(1+\delta_{in}\right)\right.\\
                &&\left.- \sum_{m>n,n=1}^3 p_{nm}\left(1+\delta_{in}+\delta_{im}\right)\right]\\
(D-4)D_{0000iii}&=&\frac{1}{240},\quad
(D-4)D_{0000iij} = \frac{1}{720},\quad
(D-4)D_{0000ijk} = \frac{1}{1440}\\
(D-4)D_{00000000}&=&\frac{1}{40320}\left[ -42 \sum_{m \ge n,n=0}^3 m_n^2  m_m^2
                  + 7 \sum_{k=0}^3\sum_{m>n,n=1}^3 m_k^2 p_{nm}\left(1+\delta_{kn}+\delta_{km}\right)
                  \right.\\
                &&\left. -2 \sum_{m \ge n,n=1}^3 p_n^2  p_m^2
                - \sum_{l>k,k=1}^3 \sum_{m>n,n=1}^3 p_{kl} p_{nm}\left(1+\delta_{kn}\delta_{lm}\right)
                 + 14  m_0^2 \sum_{n=1}^3 p_n^2 \right.\\
                &&\left.
                  - \sum_{k=1}^3\sum_{m>n,n=1}^3 p_k^2 p_{nm}\left(1+\delta_{kn}+\delta_{km}\right)
                  + 7 \sum_{m,n=1}^3 p_n^2  m_m^2 \left(1+\delta_{nm}\right)\right]\\
(D-4)D_{000000ii}&=&\frac{1}{10080}\left[ -7 \sum_{n=0}^3 m_n^2 \left(1+2\delta_{in}\right)
                 + \sum_{n=1}^3 p_n^2 \left(1+2\delta_{in}\right)\right.\\
                &&\left.+ \sum_{m>n,n=1}^3 p_{nm}\left(1+2\delta_{in}+2\delta_{im}\right)\right]\\
(D-4)D_{000000ij}&=&\frac{1}{20160}\left[ -7 \sum_{n=0}^3 m_n^2 \left(1+\delta_{in}+\delta_{jn}\right)
                 + \sum_{n=1}^3 p_n^2 \left(1+ \delta_{in}+\delta_{jn}\right)\right.\\
                &&\left.+ 2\sum_{m>n,n=1}^3 p_{nm}\left(1+\delta_{in}\delta_{jm}\right)\right]\\
(D-4)D_{0000iiii}&=& - \frac{1}{420},\quad
(D-4)D_{0000iiij} =  - \frac{1}{1680},\quad
(D-4)D_{0000iijj} =  -  \frac{1}{2520}\\
(D-4)D_{0000iijk}&=&  -  \frac{1}{5040}
\end{eqnarray*}

\section{UV-divergent parts of E-functions and F-functions }

UV-divergent expressions occur for $n \geq 3$ (E-functions) and $n \geq 4$ (F-functions) respectively, $p_{ij}$ denotes $(p_i-p_j)^2$. 

\begin{eqnarray*}
(D-4)E_{000000}&=&-\frac{1}{96} ,\quad
(D-4)E_{000000i}=\frac{1}{480}\\
(D-4)E_{00000000}&=&\frac{1}{5760}\left[ -6 m_0^2 + \sum_{n=1}^4 (p_n^2-6 m_n^2)
                  + \sum_{m>n,n=1}^4 p_{nm}\right]\\
(D-4)E_{000000ii}&=&-\frac{1}{1440} ,\quad
(D-4)E_{000000ij} = -\frac{1}{2880}  \\
\end{eqnarray*}

\begin{eqnarray*}
(D-4)F_{00000000}&=&-\frac{1}{960} ,\quad (D-4)F_{00000000i} = \frac{1}{5760}\\
(D-4)F_{0000000000}&=&\frac{1}{80640}\left[ -7 m_0^2 + \sum_{n=1}^5 (p_n^2-7 m_n^2)
                  + \sum_{m>n,n=1}^5 p_{nm}\right]\\
(D-4)F_{00000000ii}&=&-\frac{1}{20160} ,\quad
(D-4)F_{00000000ij} = -\frac{1}{40320}  \\
\end{eqnarray*}
\end{appendix}

\end{document}